# News-sentiment networks as a company risk indicator

*Thomas Forss and Peter Sarlin*

To understand the relationship between news sentiment and company stock price movements, and to better understand connectivity among companies, we define an algorithm for measuring sentiment-based network risk. The algorithm ranks companies in networks of co-occurrences and measures sentiment-based risk by calculating both individual risks and aggregated network risks. We extract relative sentiment for companies to get a measure of individual company risk and input it into our risk model together with co-occurrences of companies extracted from news on a quarterly basis. We can show that the highest quarterly risk value outputted by our risk model is correlated to a higher chance of stock price decline up to 70 days after a quarterly risk measurement. Our results show that the highest difference in the probability of stock price decline is found during the interval from 21 to 30 days after a quarterly measurement. The highest average probability of company stock price decline is seen 28 days after a company has reached the maximum risk value using our model, with a 13 percentage points increased chance of stock price decline.

Keywords: Sentiment risk; Co-occurrence; Stock price; Network analysis; Price indicator; News prediction

1. **Introduction**

If we read through different financial news sources, we are often presented with contradicting opinions and contradicting conclusions. If a person doesn't follow some kind of systematic and measurable approach when analysing a company, the person is just as likely to draw incorrect conclusions as correct ones. That is why many investors, money managers, and risk managers look at different types of indicators, ratios, and rankings to help them make decisions. To provide industry experts with a new measure, we introduce a sentiment-based company risk indicator.

Companies can be measured and ranked in many ways. The Standard & Poor's 500 (S&P 500) index is itself a measure of importance as it is an index that reflects the performance of the most influential companies in the United States stock market. One of the most widely used measures of company value is market capitalization, which is the total value of shares outstanding of a publicly traded company. Market capitalization can be used by money managers and funds as a limitation that says which companies they are allowed to invest in. Researchers have shown that investment returns are affected by firm size, however, shocks have since the 80s negatively affected smaller companies (Hou and Van Dijk 2014). Market capitalization can also be combined with other measures, such as for example book value of a company, to indicate company risk (Fama and French 1993). Traditional measures to value companies, such as for example discounted cash flows analysis, earnings ratios, price-to-book ratio, and enterprise value to EBITDA are good ways of finding relative values of companies (Damodaran 2012). However, they don't clearly state when a manager should enter, exit, or rebalance positions especially in situations when the market at large is either overvalued or undervalued. We propose that risks extracted from news could be used to support the traditional company valuation measures by adding a timing dimension.

Analysis of numerical financial data, such as stock market price movements, have stood at the centre of much of the quantitative research in finance that was done up until the early 2000s. Methods that have been used range from exploratory analysis (Capon et al. 1990) and statistical methods (Altman 1968; Pettenuzzo et al. 2013) to artificial intelligence and machine learning algorithms (Keim and Stambaugh 1986, Trippi and Turban 1992, Back et al. 1996).

Around the early 2000s researchers started showing an increased interest in financial news and this kind of research also started to gain more attention in media. That lead to among other things new types of risk assessments (Sarlin and Peltonen 2013) and predictive text-based methods (Bollen et al. 2011). Several different quantitative approaches used to measure risks have been developed since the start of the millennium (Vose 2008). Many of these new approaches have been measuring either cyclical risks (Jokipii and Milne 2008, Altunbas et al. 2010) or cross sectional risks (Acemoglu et al. 2015, Acharya et al. 2017). More recently Mezei and Sarlin (2017) put forward RiskRank, a general-purpose risk measure based on the Choquet integral that uses network theory to combine node risk with link risk in the form of indirect and direct networks. Much of the risk research so far has focused on countries, banks, and economic systems. In our research, we bring text-based risk models to corporate finance by applying RiskRank and measuring network risks that is transferred between companies found in the Standard & Poor's 500 index.

The paper is structured as follows. In section 2, we describe the text data that was gathered, and how the data contains some unique features. In section 3, we describe the models that we use. In section 4, we analyse the sentiment risk model and show how high risk equals higher risk of stock price decrease. In section 5, we summarize our findings and discuss the possible avenues of future research.

2. Data

The news articles used to test our methods were gathered as part of the research. While there are other available financial data sets online, such as Reuters data set RCV1 (Lewis et al. 2004), these types of data sets are generally only labelled according to news category, and as such is of limited use in this kind of research. Furthermore, the data sources for such news articles are quite heterogeneous as they cover any type of news that happen to fit within relatively loosely defined categories.

The data we gathered consists of crowd sourced financial news articles. These are articles that have to pass an editorial check, which is a control that they follow certain guidelines. Other than that, the articles can be written by anyone, regardless of background and intent. This type of financial news has not yet been studied to the same extent as the traditional financial news, although some research into crowd sourced data has been done (Wang et al. 2015; Zhao and Zhu 2014). Crowd sourced sites have steadily gained in popularity among investors and users. For our research, we gathered from SeekingAlpha (2017) a set of articles, which contain a body of text and the author's self-reported sentiment regarding the targeted entities in the articles. The sentiment in the articles, can be seen as a reflection of the author's future expectations.

The data gathered, consists of 17398 financial news articles starting from 2006 to the end of the second quarter in 2016, however, we limit our analysis to the period of 2011 to 2016 Q2. The articles are written by about 3600 unique authors. The articles are split and labelled by the following sectors: technology, finance, health care, consumer goods, basic materials, and services. Each article has been labelled by the author as either positive (long) or negative (short), with roughly 25% of the articles in the data set labelled as negative and about 75% of the articles labelled as positive. The content of each article is an analysis made by the author of the article on why he or she is either positive or negative toward the targeted company, commodity, sector, index, or a combination of different components.

## 3. Material and methods

In this section, we discuss the specifics of the approach used in the research. We explain how we have chosen to process the data, which analytical methods we use to create networks, and how we choose to analyse the networks. Some early research into financial news include Jacobs and Rau (1990). However, the large bulk of research using financial news, started only after the millennium changed.

In 2008, Ötzgür et al. (2008) studied networks of people in a Reuters financial news data set using co-occurrences. Other studies have since been done to rank company co-occurrence in social networks (Jin et al. 2010), and more recently mapping banking relations in text using co-occurrences (Rönnqvist and Sarlin 2015). Another area of text analysis that also has become popular is methods for analysing sentiment, also known as opinion mining. For example, researchers have shown that investor sentiment, which can be thought of as the average investor opinion can be used to predict market movements (Bollen et al. 2011), and it has also been shown that platform and type of news have different impact and longevity (Checkley et al. 2017).

### 3.1. Parsing

In order to limit the scope of the research, we decide to only look at the companies listed in the S&P 500 index. The components of the S&P 500 index varies over time as new companies emerge and other companies go bankrupt, merge, or get acquisitioned. We choose to conduct our analysis on the components of the index that were present in the index during May 2016. There were 503 components represented in the index at that time. Some of the components can be of the same company that has several classes of stocks, for example, 'GOOG' and 'GOOGL' both represent Alphabet Inc. We represent these tickers as the same company when the parsing algorithm finds either of the matches.

To parse the articles, we create a number of regular expressions that search for either the company ticker, for instance 'AAPL', or the company name 'Apple Inc.'. Whenever we match either ticker, full name, or part of the name in an article we treat it as an occurrence of the company in that article. For instance, an article mentioning 'Goldman Sachs' would be matched even though the full name of the company 'Goldman Sachs Group Inc.', was not found. A company occurring several times in an article is given no extra value and is not recorded more than once. Entity resolution for tickers was done by extracting stock tickers of traded companies in the format of (exchange:ticker), for example "(NASDAQ:TSLA)".

When developing the parsing algorithms, there are some trade-offs that need to be considered. Due to the nature of the similarity between some company names, as well as ticker symbols having different length, we decide to be quite strict while parsing occurrences. During trial runs we concluded that we would rather miss a few true occurrences and by missing instances increase the number of false negatives, than creating false positives through identifying companies that were in fact not represented in the articles. The parsing algorithm is run on a quarterly basis, but is not limited to any specific timeframe. No links between quarters is defined other than that we look for the same set of entities in each quarter.

### 3.2. Co-occurrence, sentiment, and centrality

We continue by building co-occurrence networks from the parsed news articles on a quarterly basis, starting from Q1 2011 until the end of Q2 2016. The reason that we don't include data prior to 2011 is that the frequency of articles before that point was considerably lower and we didn't want article frequency to impact our research results.

A co-occurrence network is an undirected network where in our case each company entity represents a node in the network. The links between nodes in the networks, also known as the edges, are represented by the number of occurrences the two companies have had together in articles during a specific quarter. A co-occurrence is defined as two company entities being matched in the same article. Co-mentions in the articles are of a simple nature. We look at the effects of negative and positive news for any S&P 500 companies mentioned together. This can cover partnerships, joint ventures, competitors, and suppliers, but can also cover other type of relationships, such as for example an author listing his or her opinion on current best stock picks or current worst stock picks.

A company without links to other companies, which means that the company is never mentioned together with another company in a text during the quarter, would have no edges in the co-occurrence networks. A company that frequently appears in financial news would probably have connections to other companies that not necessarily can be seen simply from reading a specific article. Two seemingly unrelated companies can, for instance, be related to each other by both having links to a third company.

As we have gathered a sentiment value for each article, we are able to distinguish between positive and negative co-occurrences. All occurrences in an article with positive sentiment are counted as positive occurrences and vice versa for negative occurrences. In the data we gathered, there is no neutral sentiment classification. To get a neutral representation, we instead combine both positive and negative occurrences into a mixed network as in Forss and Sarlin (2016). To give us a better understanding of the difference between occurrences sentiment polarizations we build three networks for each quarter. One consisting of all positive occurrences, one containing all negative occurrences, and one containing a combination of positive plus negative occurrences.

We label the network that combines all instances, as the mixed network. The relationships and links between companies can be of different nature, however, in this paper we don't have data to go to a more granular level than that which the authors themselves consider a negative or positive article. The assumption we make based on this, which also the quantitative results seem to be corroborating, is that a high percentage of negative co-mentions will in aggregate signal a higher chance of stock price decline than a high percentage of positive co-mentions. This seems to suggest that aggregating co-mentions over a period of time can, at least in the dataset that we are using, nullify the effect misinterpreted relationships.

Flow in our networks refer to how different nodes are connected to each other. Thus, the networks consist of weighted undirected edges. With more data and more advanced language parsing, such as Syntaxnet (Andor et al., 2016), it would be possible to use the semantic links linking words to extract directed networks. That could be particularly useful if it allowed us to identify pairs of occurrences where the sentiment differs depending on directionality and strength. Weighted edges between nodes in a network is generally a measure of how information spreads in the network. In the typical examples, such as the shortest path problem (Newman 2001), an edge with a low value means the nodes are close to each other. In our case the inverse is true, the higher value an edge has the more information can travel through it. However, we are interested in a measure that is able to account for all possible paths of news flow between all nodes in the networks, not only the shortest path.

Generally, different types of centrality measures are used to compare flow between nodes in a network. The relevant research on centrality in networks consists of: Identifying the most influential people in social networks (Stephenson and Zelen 1989,

Özgür et al. 2008), identifying how companies and their market cap is related in social networks (Jin et al. 2009), and identifying how banks are related in text (Rönnqvist and Sarlin 2014).

Degree centrality is one of the simplest centrality measures which uses the number of edges that a node has to measure centrality. Between-ness centrality is a centrality measure that represents how nodes in the shortest path between two nodes are connected. Eigenvector centrality is an influence measure that assumes some nodes contribute more to the measure than others, Google's PageRank algorithm is a version of eigenvector centrality. Closeness centrality calculates the distance between nodes through the shortest path to determine which nodes are important. Information centrality is a version of closeness centrality that uses harmonic mean of resistance in a graph to calculate importance of nodes. Information Centrality is one of the measures that allows us to measure flow through all paths and is the measure that we use for analysis at this stage. We start by ranking the entities in our networks according to information centrality. We use the same equation for calculating the centrality as in (Rönnqvist and Sarlin 2015) with minor modifications to fit the different types of networks.

Nodes in the network are represented by *n*. In the mixed network in equation (1) we take all the occurrences. For the long network in equation (2) we count only the positive occurrences, and for the short network in equation (3) we count only negative occurrences:

$$n = n_{positive} \qquad (1)$$

$$n = n_{negative} \qquad (2)$$

$$n = n_{negative} + n_{positive} \qquad (3)$$

Information centrality $I$ for the networks is then calculated as in (Rönnqvist and Sarlin 2015), where $i$ and $j$ are the company nodes in the networks:

$$I(i) = \frac{n}{nC_{ij} + \sum_{j=1}^{n} C_{jj} - 2\sum_{j=1}^{n} C_{ij}} \quad (4)$$

In the pseudo-adjacency matrix, $w$ is the link weight between nodes and $S(i)$ is the node weight. That gives us the following equation (Rönnqvist and Sarlin 2015):

$$C = B^{-1}, B_{ij} = \begin{cases} 1 + S(i), & if\ i = j \\ 1 - w_{ij}, & otherwise \end{cases} \quad (5)$$

To be able to rank company entities relative to each other we need to be able to calculate information centrality for all nodes in the network as one network, which is not always the case. As a side effect of having a large number of firms there is a real possibility that we will in each quarter have more than one network that are not directly linked to each other. To prevent that from happening we add Laplace smoothing as proposed by (Chen and Goodman 1996) as links between all nodes in the networks. That allows us to connect all nodes and at the same time it reduces the effects of false negatives on the network. A lower smoothing value, such as a value less than 0.5, allows the network to keep more of its characteristics, while a higher smoothing value would be mostly useful in comparing relative importance. We choose to test two different smoothing values of 0.1 and 1.0 to see if different values have a significant effect on the relative ranking of companies.

In order to get a better understanding of what the ranking represents we normalize the flow in the networks by market capitalization, and we normalize information centrality to be between 0 and 1 as the centrality values are not linear. As the market capitalization of entities varies over time we decide to use the market capitalization at the end of each quarter for normalizations. A normalized ranking should represent a ranking of companies with proportionally the highest news flow. The non-normalized ranking can be seen as representing an absolute ranking, where companies with more news in each quarter places higher in the ranking. The normalized ranking can be described through the following equation (Forss and Sarlin 2016):

$$J(i) = \left(\frac{I(i) - I_{min}(i)}{I_{max}(i) - I_{min}(i)}\right) * \frac{1}{m} \qquad (6)$$

Where $J(i)$ is the normalized information centrality of node $i$, and $m$ the market capitalization for the node at the end of that specific period. $I_{min}$ is the minimum information centrality value among values for that quarter, and $I_{max}$ is the maximum information centrality value calculated for the quarter.

### 3.3. Company risk models

Researchers have for a long time known about the common risk factors that affect returns of stocks and bonds. Fama and French (1993) introduced five common risk factors: Three for stock return and two bond risk factors that indirectly affect stock returns. These risk factors are: an overall market risk factor, risk factors that are dependent on firm size, book-to-market equity risk, bond maturity risk factors, and bond default risk factors.

Other researchers have later identified other risk factors (Whited and Wu 2006, Lando 2009). Furthermore, researchers in text analytics have been able to show that sentiment extracted from different types of sources can be used to predict stock market movements (Bollen et al. 2011, Mao et al. 2011, Hafez and Xie 2012, Checkley et al.

2017). We are interested in knowing whether news sentiment can be turned into a measurable risk indicator on a company level.

While the information centrality measure previously used to rank companies in (Forss and Sarlin 2016) showed some usefulness in analysing news flow, it didn't present the opportunity to directly analyse the effects of news sentiment on stock price movements. RiskRank on the other hand is a general-purpose algorithm for calculating an aggregate of individual network node risks and systemic risk. Mezei and Sarlin (2017) used it to calculate individual country risk as well as aggregated risks for financial systems in European countries. Crisis events were calculated using the IMF database (IMF 2017).

RiskRank provides a measure of systemic risk, as well as a measure of vulnerability of individual components. Aggregated risk using the model is calculated as the product of the individual risk multiplied with interconnected risk of components linked together as a directed graph. The aggregation is done using a discrete Choquet-integral (Labreuche and Grabisch 2003), instead of more simplistic approaches such as for example degree centrality. The Choquet-integral is here used due to the possible spill over or contagion effect between directly and indirectly connected nodes. The Choquet-integral allows for considering both interdependencies and non-linear behaviour.

The Choquet-integral can be combined with the Shapley index (Tarashev et al. 2010) when restricting subsets to a cardinality of 2. When doing that the mathematical equation used to calculate the risk can be interpreted as three different components: A direct risk value from components the node is directly connected to, an indirect value that is the spill over risk from indirectly connected nodes, and the own risk values of the node we are examining in the network. The equation can be represented as follows: (Mezei and Sarlin 2017)

$$RR(x_1, \ldots, x_n, x_c) = r_{own} + r_{direct} + r_{indirect} \qquad (7)$$

Where $r$ represents the different risk measures and $x$ are the components in the network. When adding the three different components together, we get the total risk value, which is a value with an upper bound of 1 and a lower bound of 0. The output is then an aggregated risk value between 0 and 1 for each node in the network, as well as a measure for the central node. We are not interested in the central node, which in the RiskRank algorithm is a parent node to all the other nodes. The central node is needed to be able to calculate the spill over effects between different layers in the model and allows calculating market wide effects. The three components are calculated as follows, where $c$ is the central node and $x_c$ is the node value: (Mezei and Sarlin 2017)

$$r_{own} = v(c)x_c \qquad (8)$$

$$r_{direct} = \sum_{i=1}^{n}(v(c_i) - \frac{1}{2}\sum_{j \neq i} I(c_i, c_j))x_i \qquad (9)$$

$$r_{indirect} = \sum_{i}^{n}\sum_{j \neq i}^{n} I(c_i, c_j) \prod(x_i, x_j) \qquad (10)$$

The general-purpose nature of the algorithm means that the type of risk measured reflects the type of indicators that are used as input. In most of the research on sentiment that has been published, the focus has been on market-wide investor sentiment. However, as we develop company rankings, we are interested in looking at sentiment risk on more granular level, which means looking at company specific news sentiment. In order to do that, we need to be able to do two things: We need to extract an individual sentiment measure for each company for each time period as well as a general market wide sentiment measure.

To be able to extract an individual company sentiment, we use our previously developed mixed co-occurrence matrices. From these matrices, we calculate a relative sentiment value for each company each quarter over the whole period from 2011 to 2016Q2. We define relative news sentiment for a company in any given quarter as follows:

$$s_{rel} = \frac{s_{negative}}{s_{negative} + s_{positive}} \tag{11}$$

Where $s_{negative}$ is the number of company occurrences in negative news articles and $s_{positive}$ is the number of company occurrences in positive articles for that given period. We use the relative sentiment value of each company as input to the individual risk component.

Here we identify two shortcomings in our risk model. First, the risk algorithm is not optimized for networks in the size of the S&P 500 index. That means that we need to impose limitations on the number of components that we analyse. Second, as our dataset consists of roughly 17400 articles, there is a chance that some of the less known companies will have a great fluctuation in relative sentiment from quarter to quarter. If we had an order of magnitude larger dataset, we could also do our analysis on a more granular level, for example weekly or monthly. To combat the performance problem, we use the two news flow rankings previously developed, to determine which components from the S&P 500 that are to be included in our risk analysis. After some testing, we limit our risk measure calculations to the top 50 ranking companies in both the absolute and the normalized ranking.

### 4. Results and Discussion

Here we are going to discuss the results and analyse the company co-occurrence networks and rankings both quantitatively and qualitatively. First, we do an analysis to better understand the networks and what the rankings represent.

As we have too many nodes in our networks, visualizing all 500 firms would not be understandable, because of that we will limit the visualization scope of the co-occurrence networks to the top 25 companies each quarter in coming figures. This focuses our scope enough to take a closer look at which companies are regularly present in the top 25 positive, negative, and mixed co-occurrence networks. Furthermore, in the network visualizations, we colour nodes by sectors of interest. We define the width of the edges in the visualizations of networks according to number of occurrences, and set the size of the nodes according to the total sum of the occurrences for the company in that quarter. This helps us illustrate how both individual companies and entire sectors are affected by the changes in news flow.

For the sentiment-based risk measure visualizations, we use daily stock price data as comparisons to sentiment-based risk. It is done by splitting visualizations into two horizontal subplots with the risk measure output in one plot and the stock price data in the other. The time period is then synchronized on the x-axis for both sub plots. For the risk measure, we plot all three components of risk (own, direct, and indirect) additively to get a better understanding of how the values look aggregated and the individual magnitude of each one of the three risk components.

To the left in table 1, we can see the companies that consistently can be found among the top 25 most central nodes, for each of the three different network types: positive, mixed, and negative. Apple Inc. is the only company that appears as top ranked in the positive absolute network. Bank of America Corp appears once and Apple Inc. the rest 21 quarters at the top of the mixed absolute rankings. In the negative absolute ranking we have Amazon.com, Inc. appearing four times as the most central. Netflix Inc. and Alphabet Inc., both appear as the highest ranked one time each. The rest of the 16 negative quarters Apple Inc. again is represented as the highest ranked company.

To be able to better understand the data, we test Laplace smoothing of both 0.1 and 1, and find that the changes in the ranking when increasing smoothing is marginal. Because of that, we stick to Laplace smoothing of 0.1 for the rest of the experiments. From the order of the rankings in table 1, we observe a number of patterns that could be of interest. We can look at the order of the companies that consistently appear among the highest-ranking nodes. We can also follow individual entities that change in rank over time to see a general shift in information centrality for that entity. In case the entities appearing in the top ranking barely move in terms of rankings, one could instead follow specific companies that fall or gain in rank between quarters. Third, we can compare companies between the different networks built on different sentiment, to see which companies are found in one but not the other.

The networks consist of 500 components. The average highest degree per quarter is 35, max degree for one node in the whole dataset was 88 (Apple Inc.). We calculate 1.76 average edges for the absolute mixed networks, 1.64 for the absolute positive networks, and 1.49 for the absolute negative absolute networks. We conclude that the networks are small-world networks as defined in (Floyd 1962). Furthermore, we find that the frequency of the articles grows both with time and as volatility in the markets go up.

| | Average rank top 25 | | | | | |
|---|---|---|---|---|---|---|
| | **Absolute ranking** | | | **Normalized Ranking** | | |
| Pos. | *Positive* | *Mixed* | *Negative* | *Positive* | *Mixed* | *Negative* |
| 1 | AAPL | AAPL | AAPL | NFLX | NFLX | NFLX |
| 2 | MSFT | GOOG | AMZN | NVDA | BBY | BBY |
| 3 | GOOG | MSFT | GOOG | BBY | NVDA | NVDA |
| 4 | AMZN | AMZN | MSFT | IP | IP | YHOO |
| 5 | IBM | WMT | NFLX | QRVO | YHOO | IP |
| 6 | WMT | NFLX | IBM | GPS | GPS | CMG |
| 7 | FB | IBM | INTC | DNB | DNB | DO |
| 8 | INTC | FB | FB | YHOO | SPLS | SPLS |
| 9 | BAC | INTC | WMT | JNPR | JNPR | AVGO |
| 10 | HPQ | HPQ | YHOO | FSLR | AVGO | M |
| 11 | GS | BAC | HPQ | SPLS | CMG | CRM |
| 12 | KO | GS | GS | SEE | FSLR | JNPR |
| 13 | BRK | YHOO | CSCO | AVGO | SEE | MLM |
| 14 | JPM | CSCO | CRM | DPS | QRVO | DNB |
| 15 | CSCO | MS | VZ | UA | DO | TGT |
| 16 | MS | KO | TGT | EA | UA | RIG |
| 17 | NFLX | IP | ORCL | MU | EA | KSS |
| 18 | C | JPM | CMCSA | HPQ | MU | AMZN |
| 19 | YHOO | ORCL | MCD | GT | M | GPS |
| 20 | IP | VZ | T | HRS | GT | HPQ |
| 21 | WFC | C | CMG | DO | HPQ | TWC |
| 22 | ORCL | BRK | DIS | AA | CHK | MS |
| 23 | GM | WFC | TWX | CHK | COH | COH |
| 24 | XOM | T | MS | CMG | CRM | SEE |
| 25 | JNJ | GM | IP | HRB | DPS | CA |

Table 1. List of companies appearing in the top 25 of either the absolute or normalized rankings. We provide a ranking for each of the three types of co-occurrence networks. The mixed network ranking contain all articles, where the positive ranking contains only articles that where labelled by the author as positive, and the negative ranking only contains articles that were labelled as negative by the authors.

As larger companies seemed to dominate the different absolute rankings, we also created a normalized version of the ranking. The hope was that a normalized ranking would represent companies that have a higher news flow relative to their size. To the right in table 1, we recalculate the ranking using the two normalizations that were presented in section 3 using equation (6). In table 1, to the right, we see that when re-calculating the ranking Apple falls from first place to not being in the top 25 ranking anymore. In the normalized average, Netflix Inc. places at top of the average ranking for all three types of networks.

To get a better understanding of what the changes in information centrality mean we plot information centrality of the 25 highest ranked firms from the two mixed networks. In figure 1 we plot the absolute components and in figure 2 we plot the normalized components. Apples' dominates the absolute ranking in figure 1, while the other components in the ranking often change between quarters. From figure 2 we can see that the values in the normalized ranking is closer to each other and positions in the ranking changes regularly. Some of the companies are found in both rankings, Yahoo is one example. Both rankings follow a similar general pattern with a couple of spikes in information centrality at different periods. The two largest spikes are found the middle of 2012 and at the beginning of 2016, possibly as a reaction to major events in the markets, at least the spike in 2016 can be explained by market volatility increasing. Both rankings information centrality seems to be bottom out at just above 0.05, and that can largely be attributed to the smoothing value of 0.1 that was used. As comparison, we find that the lower bound of the rankings is around 0.5 at a smoothing value of 1.

We continue our results analysis by looking at risk values of the top performing companies in both the absolute and normalized rankings. Performance constraints in the risk algorithm limit the number of components that we can include in the statistical modelling. Because of that we include the companies in our analysis that on average place in the top 50 in either the mixed absolute ranking or top 50 in the mixed normalized ranking. We choose the top performing components because they have a higher news flow than lower performing components, and should thanks to that be less susceptible to variance and volatility. 12 of the 100 companies are found in both top 50 of both rankings giving us a total of 88 unique companies to analyse. The data we use for this experiment spans 22 quarters. However, some quarterly data points are disqualified from the analysis due to lack of stock price data. After disqualifying data points we end up with a total of 1864 valid quarterly data points. When doing performance benchmarks we use the total amount of data points as the benchmark comparison.

In figure 3, we plot Apple Inc. stock price and risk values for the period. Risk is visualized as three lines, the bottom line is the individual risk, the middle line represents direct risks added to the individual risks, and the uppermost line represents all three risk components aggregated. In figure 4, we visualize the spread of quarterly data points for aggregated company risk and individual company risk at intervals of 0.1. We can see that the aggregated risk values increase the number of data points at all risk thresholds. At a risk threshold of higher than 0.4 the aggregated risk values are more than doubled compared to data points that only take into account individual risk. At a risk threshold of 1.0 the individual data points are only about 3.5% of the total data points, whereas the aggregated risk data points are roughly 9.4% of the total data points.

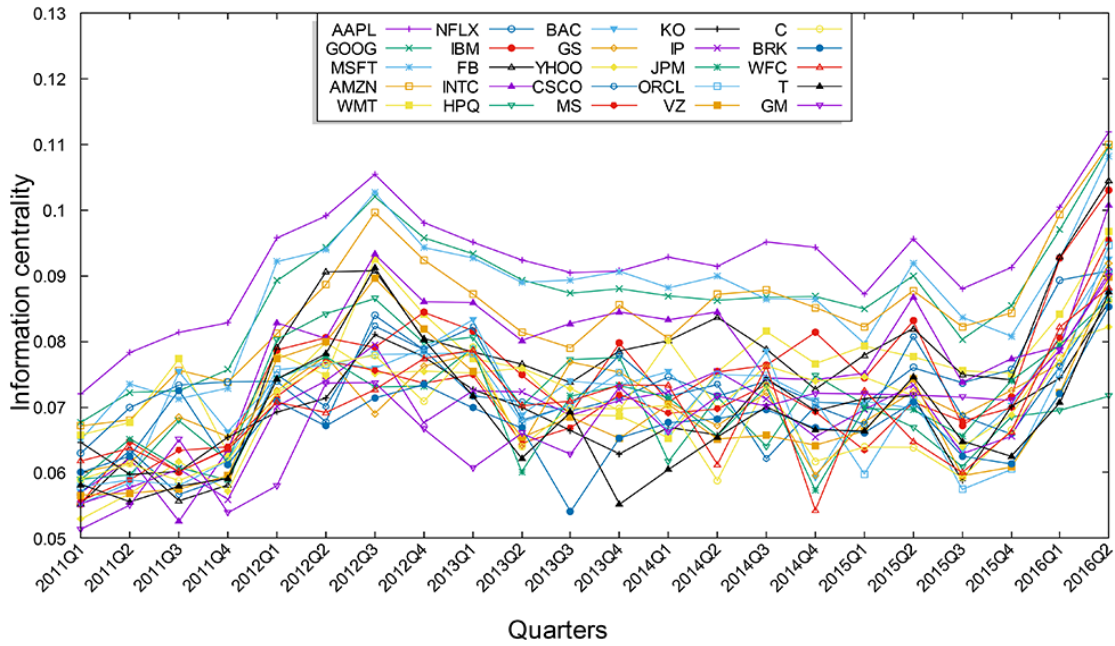

Figure 1. Line graph of information centrality for the average top 25 companies in the absolute ranking.

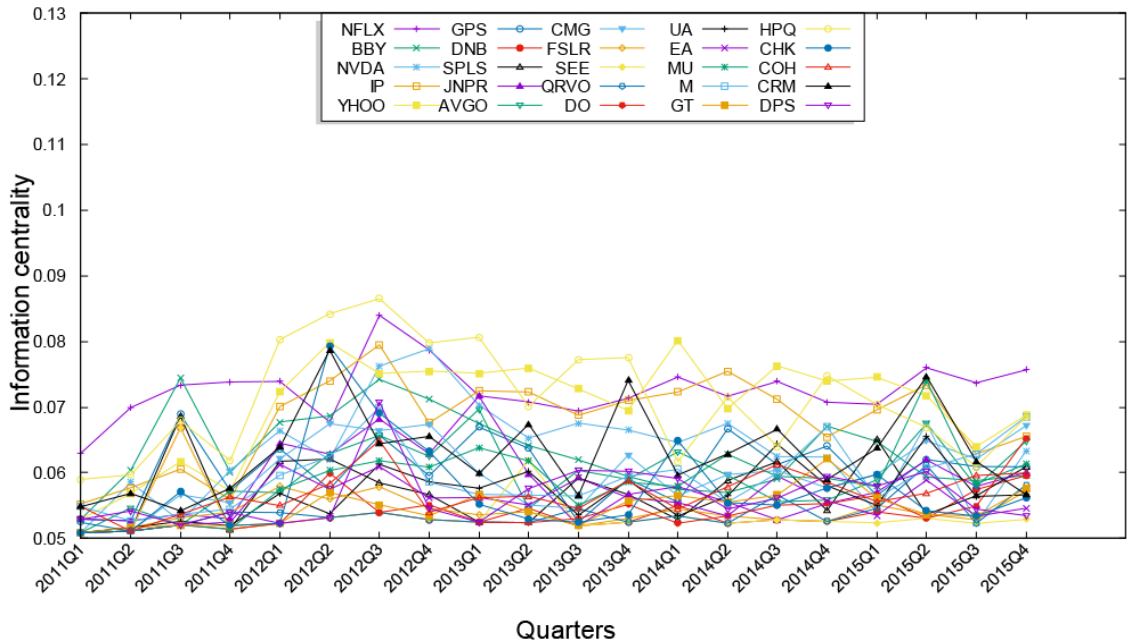

Figure 2. Line graph of information centrality for the average top 25 companies in the normalized ranking, graph ends at Q4 2015 due to lacking market capitalization data for the last two quarters.

To show the statistical significance of the risk values, we compare percentage of company stock price decreases for different subsets of data points limited by a risk threshold, against percentage of company stock price decrease of the total number of data points (the benchmark). By doing that we automatically take into account the upward long-term bias that the stock market has shown since its inception. To further analyse the usefulness of aggregating individual risk values with network risk, we also compare subsets of individual risk against subsets of aggregated risk. As the measurements are done on a quarterly basis, we set the delay of adjusted closing stock price values that we analyse, to a range between 3 and 90 days.

As the risk values that we have defined are measured between 0 and 1, we do benchmark calculations for data points with risk values between 0.5 and 1 with increments of 0.1. Lastly, we analyse stock price decrease in different ranges of days. This is done to determine if the aggregated risk values have a greater statistical impact than individual risk values at different delay intervals.

As can be seen in table 2, aggregated risk outperforms the comparison benchmark of all data points with more than two standard deviations for the period of 11 to 70 days' delay, and with over 10 standard deviations for the period between 21 to 50 days. Furthermore, from table 3 we can also see that when comparing individual risk against network risk, the network risk consistently outperforms individual risk at a risk threshold of 1 for the periods between 11 days and 50 days' delay. It is worth noting that identifying more data points is valuable on its own even in the situations where network risk isn't significantly outperforming the individual risk thresholds.

**Stock price to risk comparison**

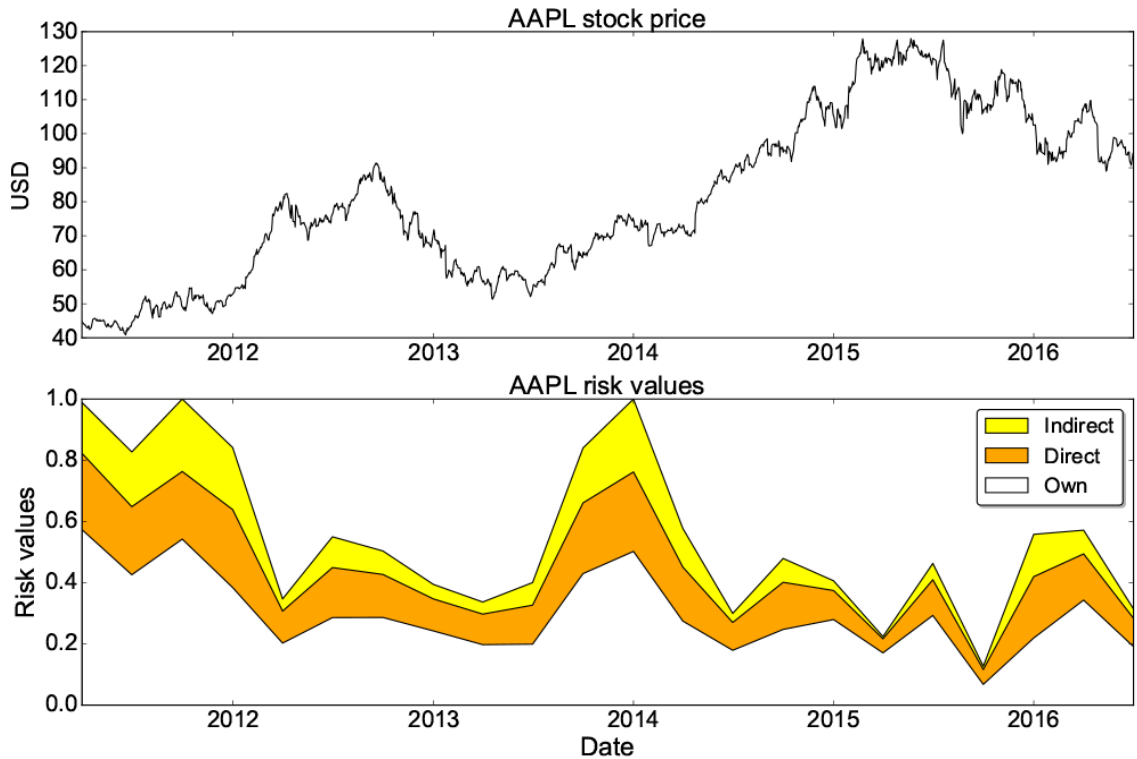

Figure 3. Side-by-side comparison of the risk measures against stock price for Apple Inc. over the period 2011 to Q2 2016.

When comparing stock price decrease for different aggregated risk thresholds against the benchmark for individual days, we find that a 28 days delay at risk threshold 1.0 has the highest difference, a difference of 13.14 percentage points (55.68% chance of decrease vs. 42.54% for the benchmark). Furthermore, we find that all days between 25 days and 30 days outperform the benchmark by over 10 percentage points at threshold 1.0. When comparing risk thresholds below 1.0, we find that for the delay range of 41 to 50 days the risk threshold 0.6 to 0.9 outperforms threshold 1.0 when compared relative to the benchmark. We also find that the individual risks at the lower thresholds have performance in line with aggregated risk.

| Days delay | Stock price decrease at risk threshold 1.0 | | | | | |
| --- | --- | --- | --- | --- | --- | --- |
| | Aggregate decrease % | Comparison decrease % | Absolute difference | Relative difference % | Standard deviation | St.dev. difference |
| **3 to 90** | 46.93 | 42.11 | 4.82 | 11.44 | 2.82 | 1.71 |
| **3 to 45** | 50.74 | 43.71 | 7.03 | 16.09 | 3.09 | 2.28 |
| **45 to 90** | 43.28 | 40.58 | 2.70 | 6.65 | 1.29 | 2.09 |
| **3 to 10** | 50.21 | 45.29 | 4.92 | 10.86 | 5.36 | 0.92 |
| **11 to 20** | 52.67 | 46.22 | 6.45 | 13.95 | 1.52 | **4.24** |
| **21 to 30** | 51.59 | 42.00 | 9.59 | 22.83 | 0.72 | **13.41** |
| **31 to 40** | 50.11 | 42.75 | 7.36 | 17.22 | 0.56 | **13.22** |
| **41 to 50** | 47.33 | 41.52 | 5.81 | 13.98 | 0.56 | **10.41** |
| **51 to 60** | 47.95 | 41.80 | 6.16 | 14.73 | 0.88 | **6.96** |
| **61 to 70** | 41.99 | 40.14 | 1.85 | 4.61 | 0.66 | **2.82** |
| **71 to 80** | 41.31 | 40.17 | 1.14 | 2.83 | 1.29 | 0.88 |
| **81 to 90** | 39.83 | 39.73 | 0.10 | 0.26 | 1.22 | 0.08 |
| **Average** | 47.00 | 42.17 | 4.83 | 11.29 | 1.66 | 4.92 |

Table 2. Comparison of stock price decrease for aggregated risk against the benchmark consisting of all analysed data points. To the right we can see the number of standard deviations that the aggregated risk decrease is above the benchmark.

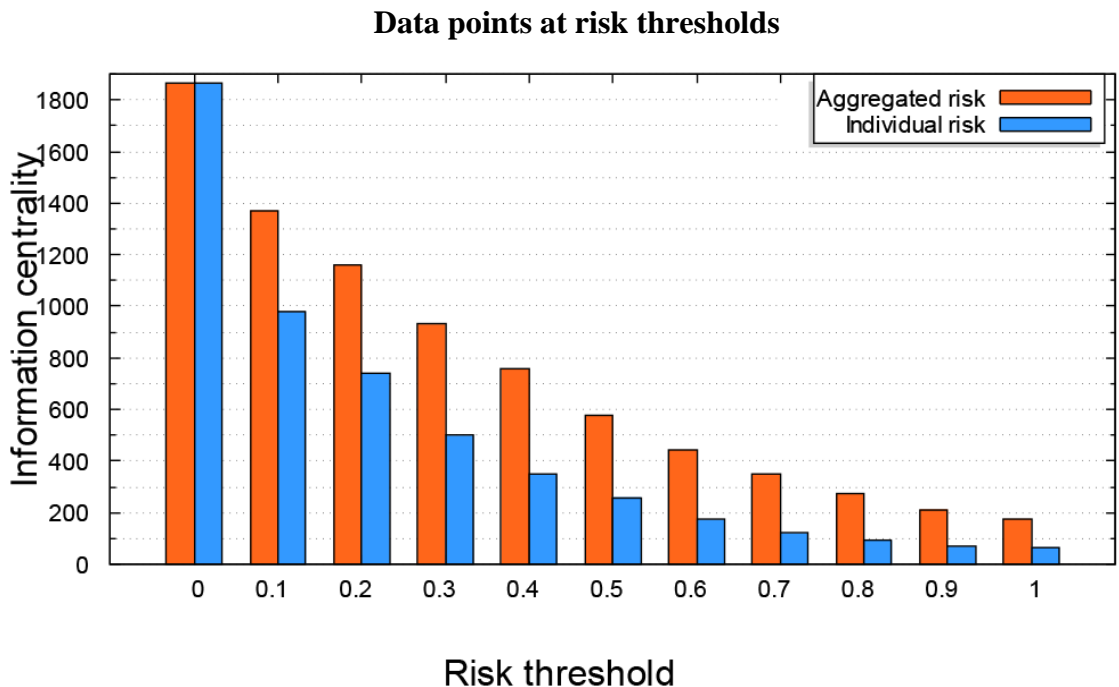

Figure 4. Comparison of available data points at different risk thresholds, aggregated vs. individual risk. 1864 total data points.

| Aggregated vs. individual stock price decrease at risk threshold 1.0 | | | | | | | | |
|---|---|---|---|---|---|---|---|---|
| Days delay | Agg. % decrease | Ind. % decrease | Comp.% decrease | Agg. diff. | Ind. diff. | Comp. st.dev. | Agg.st. dev.diff. | Ind.st. dev.diff. | Agg. outperf. |
| 3 to 90 | 46.93 | 44.11 | 42.11 | 4.82 | 2.00 | 2.82 | 1.71 | 0.71 | 1.00 |
| 3 to 45 | 50.74 | 45.42 | 43.71 | 7.03 | 1.71 | 3.09 | 2.28 | 0.55 | 1.72 |
| 45 to 90 | 43.28 | 42.86 | 40.58 | 2.70 | 2.28 | 1.29 | 2.09 | 1.77 | 0.33 |
| 3 to 10 | 50.21 | 41.86 | 45.29 | 4.92 | -3.44 | 5.36 | 0.92 | -0.64 | 1.56 |
| 11 to 20 | 52.67 | 48.33 | 46.22 | 6.45 | 2.11 | 1.52 | 4.24 | 1.39 | **2.85** |
| 21 to 30 | 51.59 | 45.91 | 42.00 | 9.59 | 3.91 | 0.72 | 13.41 | 5.46 | **7.95** |
| 31 to 40 | 50.11 | 44.85 | 42.75 | 7.36 | 2.10 | 0.56 | 13.22 | 3.77 | **9.46** |
| 41 to 50 | 47.33 | 45.91 | 41.52 | 5.81 | 4.39 | 0.56 | 10.41 | 7.87 | **2.55** |
| 51 to 60 | 47.95 | 48.03 | 41.80 | 6.16 | 6.23 | 0.88 | 6.96 | 7.05 | -0.09 |
| 61 to 70 | 41.99 | 41.06 | 40.14 | 1.85 | 0.92 | 0.66 | 2.82 | 1.40 | 1.42 |
| 71 to 80 | 41.31 | 41.52 | 40.17 | 1.14 | 1.34 | 1.29 | 0.88 | 1.04 | -0.16 |
| 81 to 90 | 39.83 | 39.09 | 39.73 | 0.10 | -0.64 | 1.22 | 0.08 | -0.52 | 0.61 |
| Average | 47.00 | 44.08 | 42.17 | 4.83 | 1.91 | 1.66 | 4.919 | 2.487 | 2.432 |

Table 3. Comparison of stock price decrease for aggregated risk against individual risk at threshold 1. To the right we can see the aggregated risk standard deviation outperformance over the benchmark subtracted by the outperformance of the individual risks over the benchmark.

Finally, we calculate the standard errors at risk threshold 1 between aggregated data points, individual data points, and the benchmark data points as they contain different sized proportions. We compare the periods of 11 to 20 days and 61 to 70 days against the benchmark as those periods have the lowest standard deviation outperformances. For the aggregated risk the probability error ends up rounded to 0.09 for both ranges, and for the individual risk that has a smaller set the probability errors are both rounded to 0.14. We calculate the standard error between aggregated risk and individual risk at threshold 1 for the ranges 11 to 50 days. The probability error ends up at 0.46 for all four ranges. Based on this we are able to conclude that our findings are statistically significant when compared to the benchmark.

## 5. Conclusion

Based on our quantitative analysis we are able to confirm that network sentiment-risk extracted from crowd sourced financial news can be used to indicate days and/or ranges of days when a company has an increased probability of stock price decrease. Previous studies where researchers have extracted sentiment from micro-blogging sources has shown that type of sentiment shows predictive power on varying timespans. Our approach that aggregates news over quarters shows statistical predictive potential between delays of 11 to 70 days. Furthermore, our results show that the range with highest predictive potential is found between delays of 21 to 50 days when doing quarterly sentiment measurements, during which aggregated network on average is more than 10 standard deviations above the benchmark. In our experiment a delay of 28 days at risk threshold of 1 showed the highest outperformance when compared to the benchmark, and a decrease difference of 13.14 percentage points.

Our findings lead us to believe that there are at least four major factors at play in determining when sentiment will show predictive potential. First, the type of financial data sources used may affect when and how predictive potential manifests. Breaking-news type of sources like for instance tweets (Checkley et al., 2017) seem to have a more immediate effect than full length articles. Second, the type of sentiment extracted seems to affect predictive potential. Extracted sentiment vs. self-reported sentiment could be a factor in how predictive potential manifests. Third, the period over which we choose to measure sentiment likely affects the results. In future research, we will look further into this by measuring sentiment in different intervals and creating rolling sentiment indicators. Fourth, the network effects that our approach is based upon could also partly explain the increased delay in predictive potential compared to other previous research in sentiment. While we did compare the individual risks vs.

aggregated risks we must keep in mind that both these approaches were based on underlying co-occurrence networks. To find out whether this affects the results, part of our future research will consist of comparing predictive power of simple company occurrences in the same dataset to the predictive power of co-occurrences.

Based on having shown that the sentiment risk holds predictive power, we can further develop this into a prescriptive risk tool. The tool could be used to warn when companies are at higher risk of stock price decrease, something that would be directly useful for finance industry experts, such as portfolio managers, investors, and traders. While the methods were only tested on a quarterly basis, we plan on extending the work to be able to provide daily risk insights by creating rolling sentiment risk values. Such a tool could be used by anyone regardless of analytical skill as the output the prescriptive system would give is companies and their associated risk values.

Finally, as we were only able to determine the likelihood of stock price decrease, and not the magnitude of the moves, we will continue our research by developing portfolio strategies and back-test the portfolio results against index benchmarks. We are looking into using artificial intelligence, such as Neural Networks, to predict magnitudes of moves based on sentiment-risk.

**Acknowledgements**

Thomas Forss would like to thank Åbo Akademi University for the support and Prof. Dr. Christer Carlsson for the feedback.